\documentclass{article}
\def\longonly#1{}
\usepackage[T1]{fontenc}
\usepackage{graphicx}
\usepackage{amsfonts}
\usepackage{hyperref}
\begin{document}
\title{Simulating, Visualizing and Playing with de Sitter and anti de Sitter spacetime}
\author{Eryk Kopczy\'nski\\ \small{Institute of Informatics, University of Warsaw}}
%
%

\def\bbE{\mathbb{E}}
\def\bbS{\mathbb{S}}
\def\bbR{\mathbb{R}}
\def\bbH{\mathbb{H}}
\def\ra{\rightarrow}
\def\acos{\arccos}
\def\acosh{\mbox{arcosh}}
\def\dS#1{d\bbS^#1}
\def\wadS#1{ad\bbS^#1}
\def\uadS#1{\widetilde{ad\bbS^#1}}

\maketitle              
\begin{abstract}
In this paper we discuss computer simulations of de Sitter and anti de Sitter spacetimes,
which are maximally symmetric, relativistic analogs of non-Euclidean geometries.
We present prototype games played in these spacetimes; such games and visualizations
can help the players gain intuition about these spacetimes. We discuss the technical challenges in creating
such simulations, and discuss the geometric and relativistic effects that can be witnessed by 
the players.
\end{abstract}
\section{Introduction}
Science-based games are games based on a real scientific phenomenon.
For example, there are
games based on special relativity \cite{SlowerSpeed,VelocityRaptor}, quantum mechanics \cite{MigdalQuantum2}, orbital physics
\cite{Kerbal},
non-Euclidean geometry \cite{weeksgames,hyperrogue,hyperbolica,weeksbilliards}. Compared to typical educational games, where the 
concept explained and the gameplay are not related, science games take the
scientific phenomenon and use it to create interesting gameplay \cite{MigdalQuantum2}. Other than
providing entertainment, science games provide intuitive unterstanding of difficult
science concepts; many of them let the player perform their own experiments,
by including sandbox elements, level editors, or being open source. Such a better
understanding helps young players consider a scientific career, and also gives ideas for
new scientific developments to mature researchers \cite{hamilton2021nogo}. Science games require a specialized 
engine for efficient modelling and visualization of the given scientific concept; such 
engines may be later used for other applications than just games \cite{hrviz}.

In this paper, we describe a game (or rather, a collection of two games) taking place in de Sitter and anti de Sitter spacetimes.
These spacetimes could be seen as relativistic analogs of spherical and hyperbolic geometry.
De Sitter spacetime is of interest as an asymptotic approximation of our universe \cite{expDeSitter}, and anti-de Sitter
spacetime is of interest for its correspondence with conformal field theory \cite{Maldacena1999}.
While many games exist based on special relativity and non-Euclidean geometry, the unintuitive
properties of spacetimes combining these two concepts seems to be still unexplored.

Players can try our prototype game, named {\it Relative Hell}, by downloading the Microsoft Windows binary\footnote{\url{https://zenorogue.itch.io/relative-hell}, 
source code: \url{https://github.com/zenorogue/hyperrogue/blob/master/rogueviz/ads/ads-game.cpp}. Last accessed April 9, 2024.}.
The source code, based on the non-Euclidean engine RogueViz \cite{rogueviz2023}, is also available.

\begin{figure}[t]
\begin{center}
\includegraphics[width=.3\linewidth]{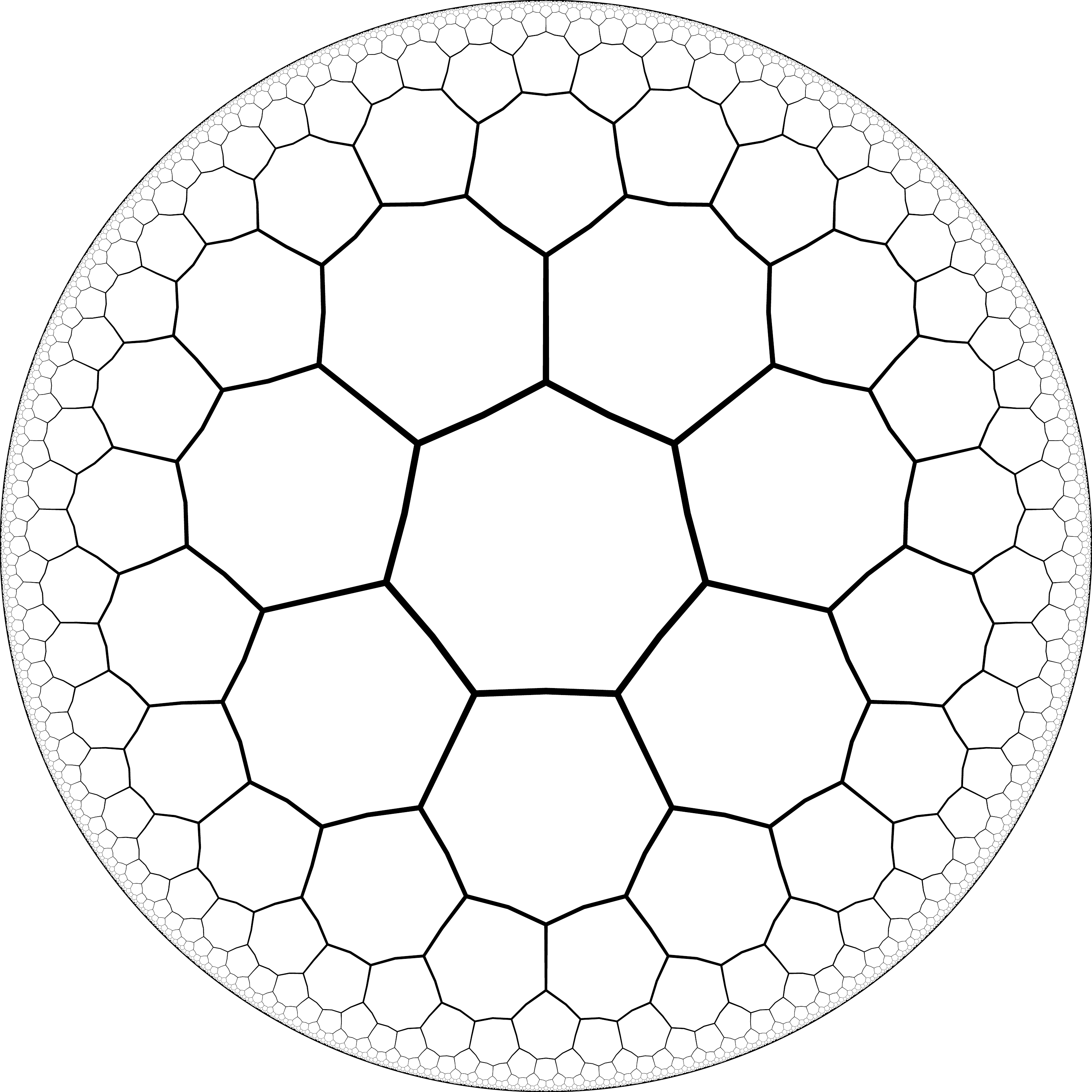}
\includegraphics[width=.3\linewidth]{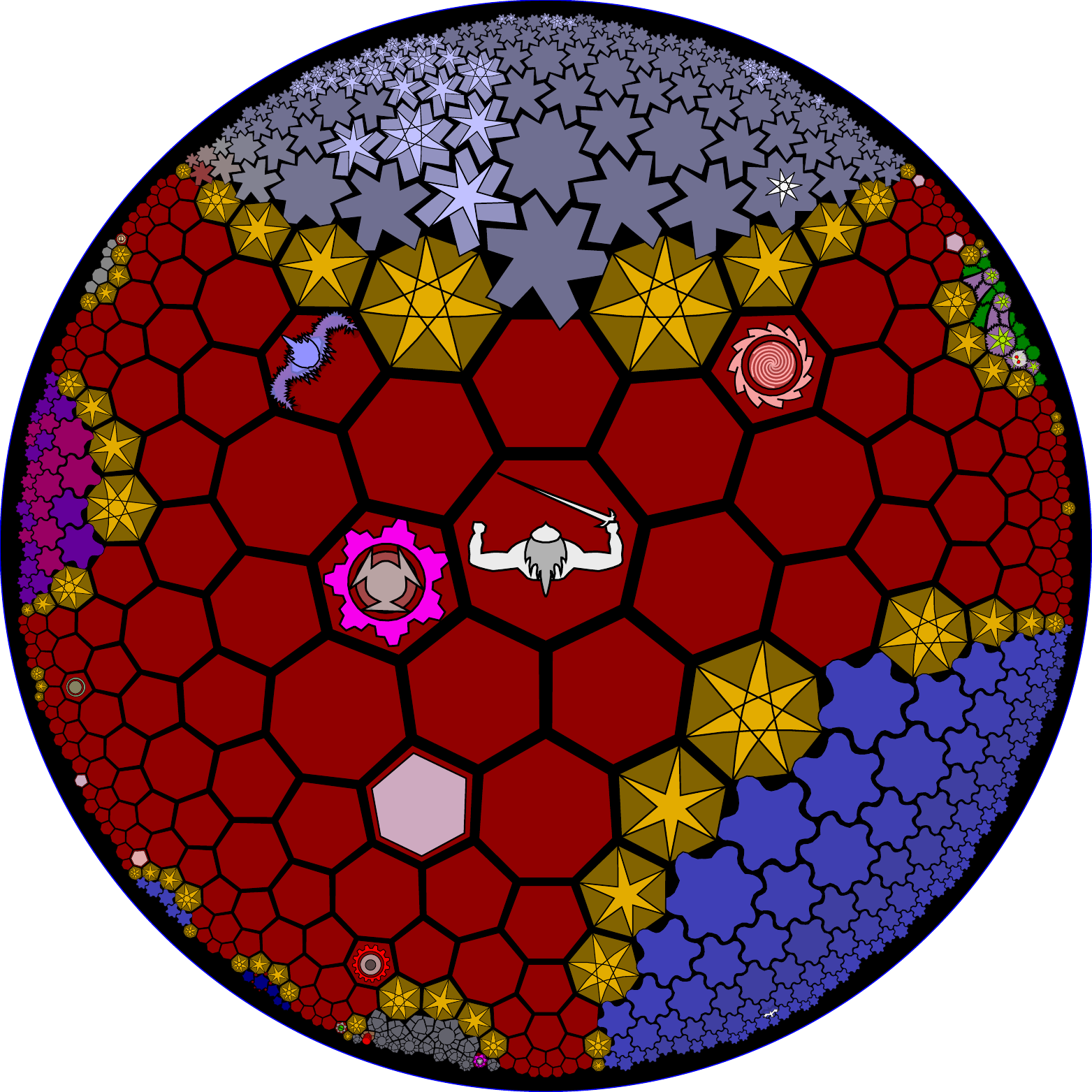}
\includegraphics[width=.3\linewidth]{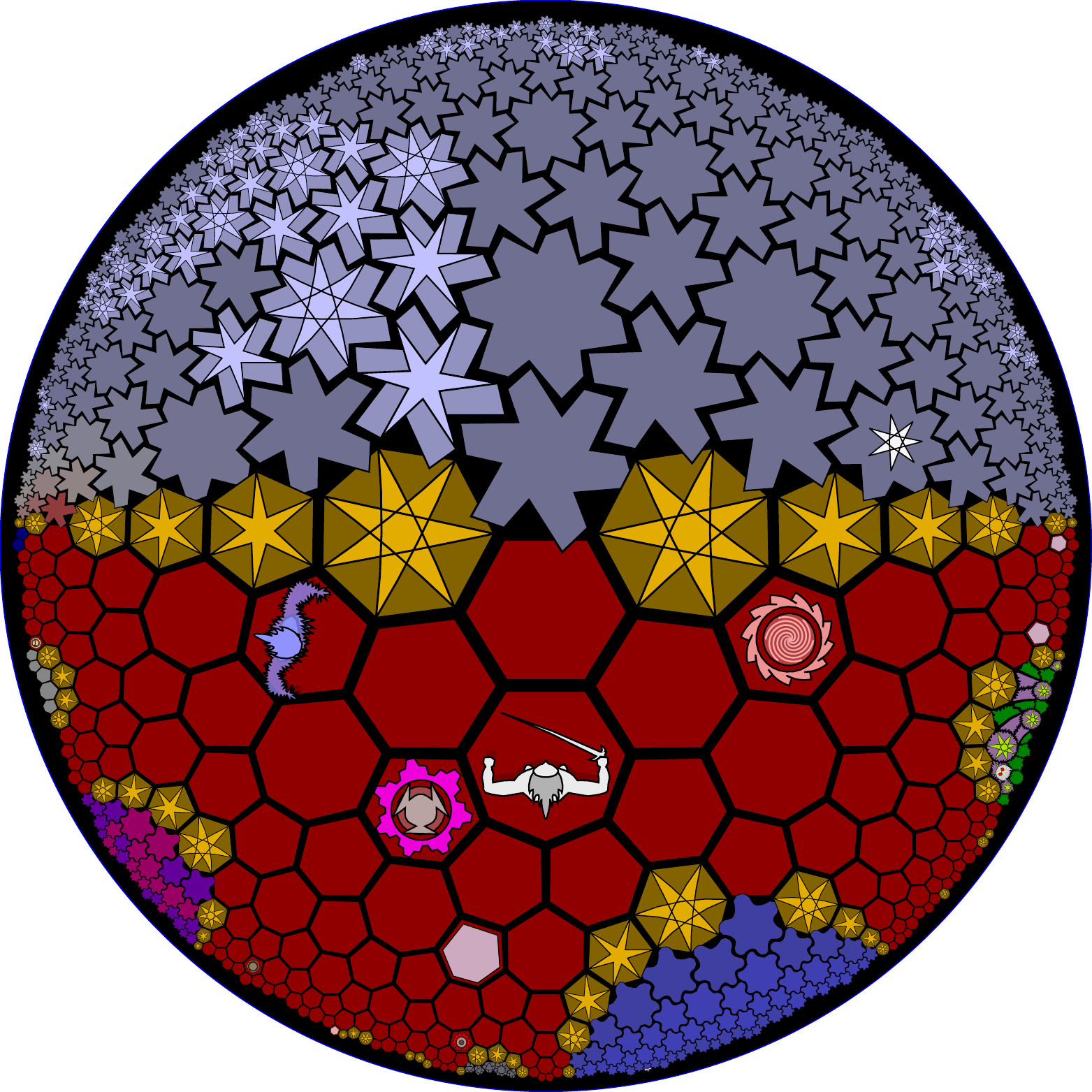}
\end{center}
\caption{On the left, the \{7,3\} hyperbolic tessellation in Poincar\'e disk model.
The Poincar\'e disk model is conformal: it does not distort small shapes, so all heptagons
look close to regular; however, it distorts scale: all the heptagons shown are of the same
size.
On the right, the same scene in \emph{HyperRogue} viewed from two points. In the Poincar\'e model,
straight lines are projected as circular arcs orthogonal to the disk boundary;
moving the center shows the player that the walls (orange) are indeed straight lines.
\label{fig73}
}
\end{figure}

\section{Background}

Our prototype helps non-experts understand scientific concepts intuitively.
In this section, we provide an intuitive introduction to non-Euclidean geometry and relativity theory,
and explain how games can provide such understanding.

The history of non-Euclidean geometry starts with Euclid's \emph{Elements}. This book has changed
teaching geometry by giving it structure: starting with very basic \emph{postulates} and \emph{axioms},
such as the space never ending and being the same everywhere and in every direction, and continuing with
all the more complex geometric facts known to Euclid, which followed from the postulates and axioms \cite{coxeterneg}. However,
some geometric facts related to \emph{parallel lines} did not actually seem to follow from such basic
postulates. For example, if we have two points $A_1B_1$ in distance $d$ on a straight line $l_1$ and two points $A_2B_2$
in distance $d$ on a parallel straight line $l_2$, the distance between $A_1$ and $A_2$ should be always the same as
between $B_1$ and $B_2$. Euclid has solved this by declaring (an equivalent formulation of) this
statement as his \emph{fifth postulate} (or \emph{parallel postulate}).

Mathematicians have been trying to prove the \emph{parallel postulate} from the other postulates, until 
Lobachevsky and Bolyai have discovered \emph{hyperbolic geometry} $\bbH^d$, which satisfied all Euclid's postulates
except his \emph{fifth postulate} \cite{coxeterneg}. (In $\bbH^d$, $d$ is the number of dimensions.) To explain how the parallel postulate could not be true,
imagine drawing great circles on a sphere; such great circles are an analog of Euclidean lines,
however, if we look at two meridians, we discover that the parallel postulate does not
hold. In our three-dimensional world, great circles are obviously curved; however, we can imagine that our
Universe is actually a hypersphere in four dimensions, and what we perceive as straight lines is actually
great circles on this hypersphere. This is the \emph{spherical geometry} $\bbS^d$; hyperbolic geometry 
is the opposite of it (while, in spherical geometry, ``parallel'' lines converge, in hyperbolic geometry
they diverge). While many everyday objects are spherical, hyperbolic geometry is more difficult to
understand; a good way to obtain intuitive understanding is to play games, aiming to simulate the
experience of a inhabitant of a non-Euclidean world. Games in $\bbH^2$ \cite{weeksgames,hyperrogue,visiccs}
typically display the view in the \emph{Poincar\'e disk model} (Figure \ref{fig73}, which is a projection of hyperbolic plane $\bbH^2$
to a Euclidean disk, centered at the player character's position; this lets the player see that straight lines
are indeed straight and acting strangely. There are also immersive visualizations of $\bbH^3$ \cite{gunnvis,weeksrealh,hyperbolicvr,hyperrogue,hyperbolica}.

While the world was originally assumed to be Euclidean, further experiments have shown this to not to be
the case. The Morley-Michelson experiment has shown that the speed of the light measured by
a moving observer is always the same (independent by the observer's velocity), which was in conflict with
theories at that time. This was resolved in the theory of special
relativity: Euclidean space and Newton's notion of time were replaced by \emph{Minkowski spacetime};
there was no absolute time -- for two observers $O_1$ and $O_2$ meeting at position 0 in time 0, an event
at position $x$ and time $t$ according to $O_1$ would be at position $x'$ and $t'$ according to $O_2$,
with both space and time being changed by Lorentz transformations ($x\neq x'$ and $t\neq t'$), a bit
similar to how spatial rotations change both $x$ and $y$ coordinates, and causing effects such as
contraction of lengths and dilation of time. Special relativity effects normally require large speed
to observe in the real world, but again, they can be experimented with in games \cite{SlowerSpeed,VelocityRaptor}.
Special relativity could not explain gravity; for this, we need general relativity: the notion of curved spacetime, similar to the curved space of non-Euclidean geometry.
Furthermore, as explained in Section \ref{ref:prelim} below, Minkowski spacetime geometry can be used to provide an elegant model of $\bbH^d$.

\begin{figure}[t]
\begin{center}
\includegraphics[width=.3\linewidth]{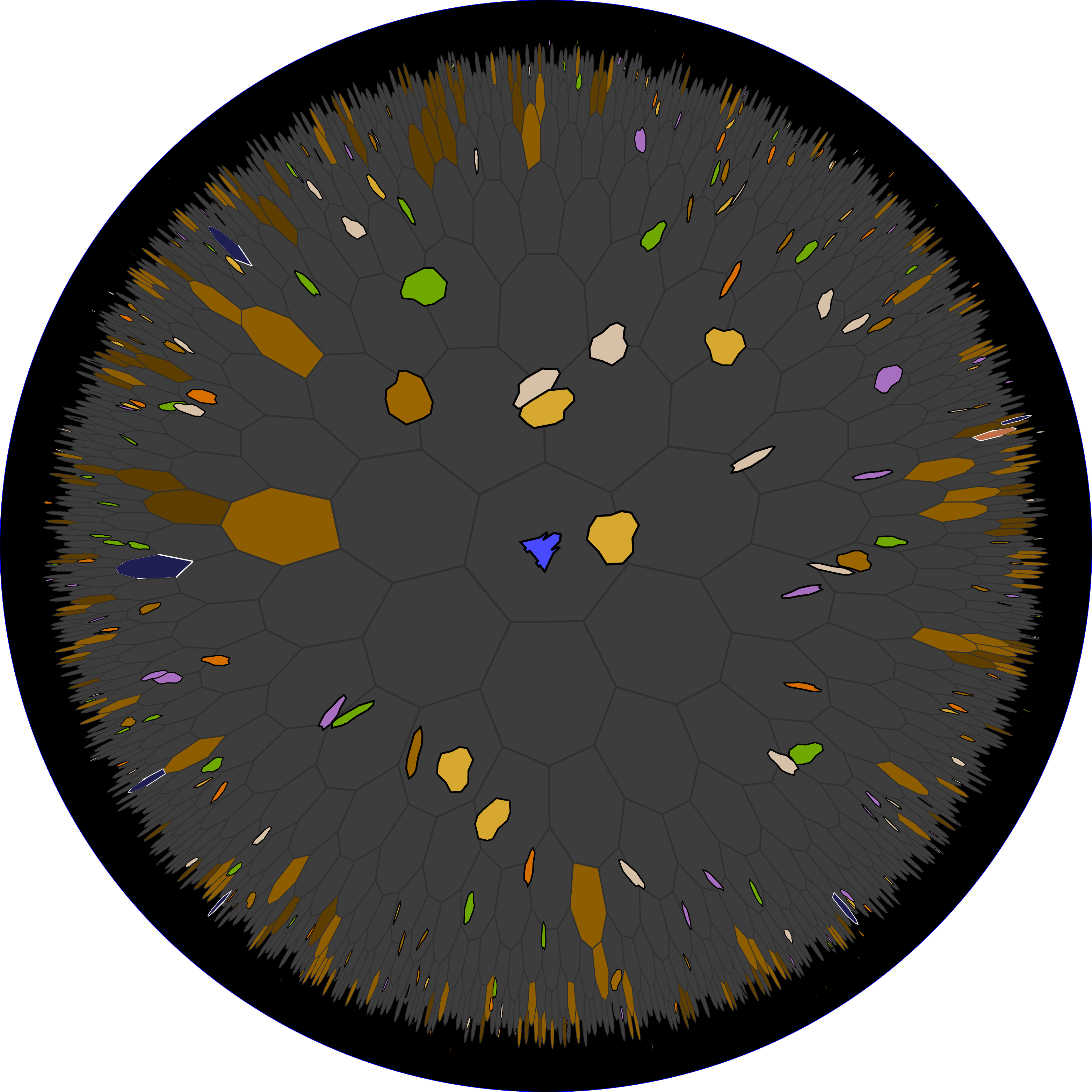}
\includegraphics[width=.3\linewidth]{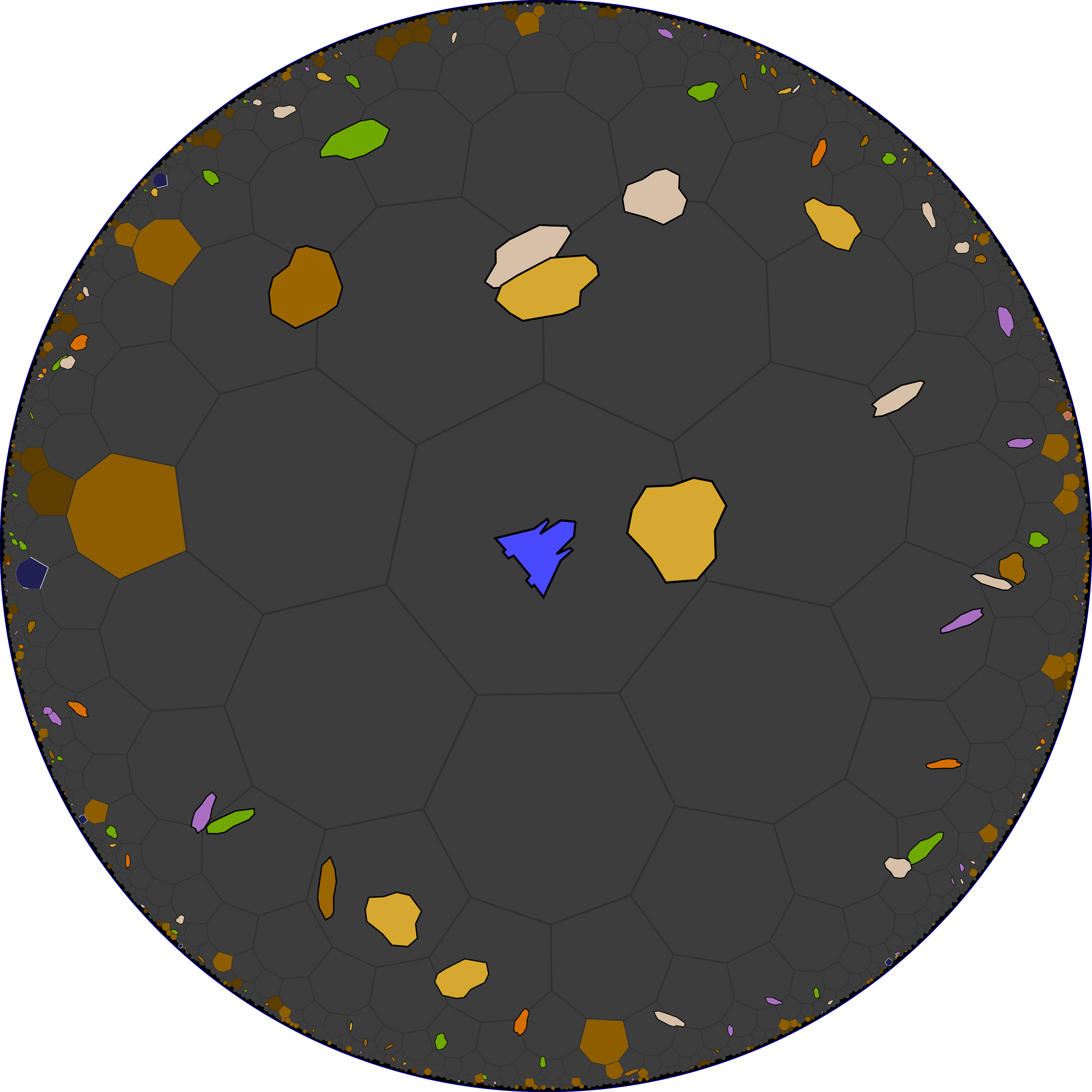}
\includegraphics[width=.3\linewidth]{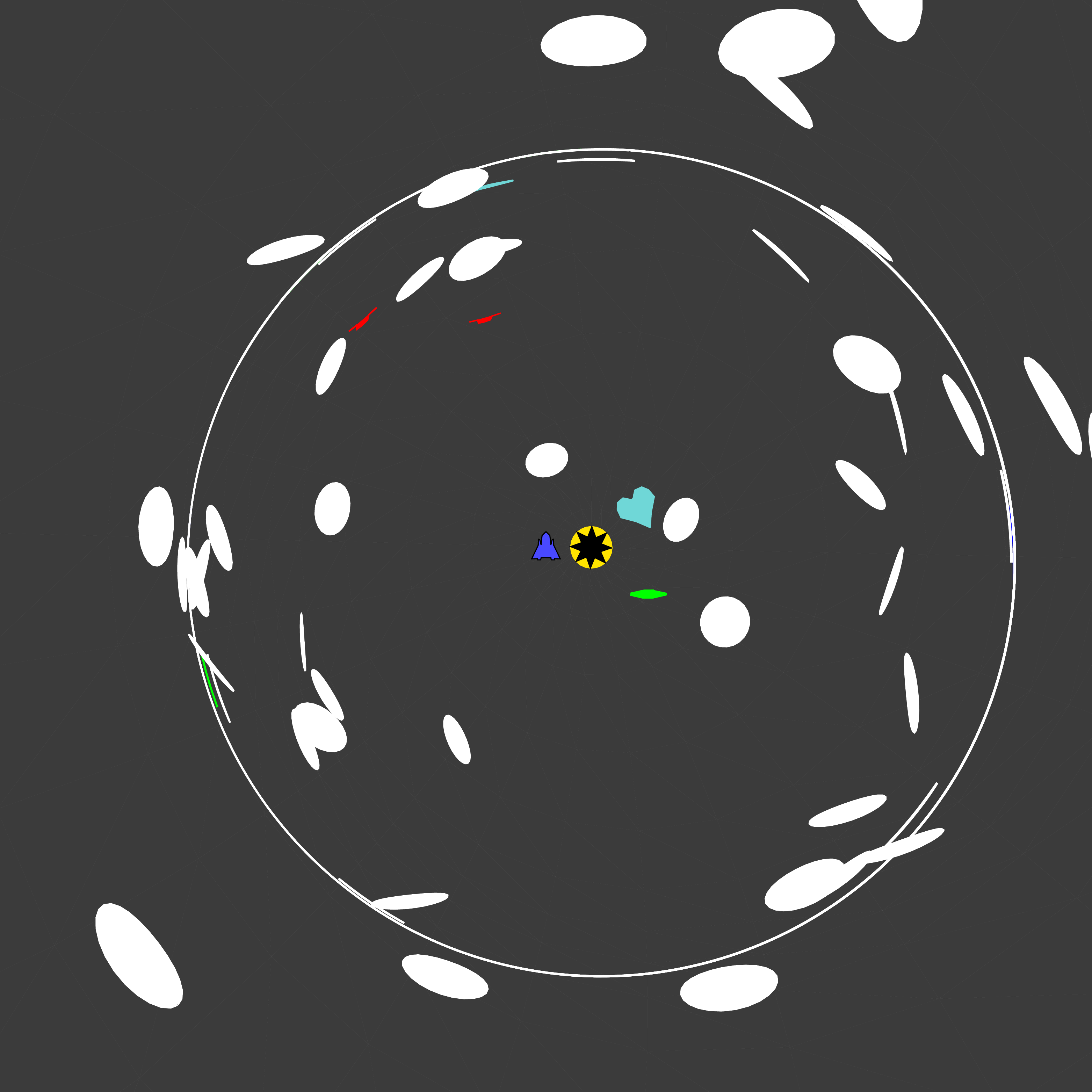}
\end{center}
\caption{Relative Hell. $\uadS{2}$ is displayed in the Poincar\'e disk model on the left, and the Beltrami-Klein disk model in the center.
$\dS{2}$ on the right, in stereographic projection.
\label{relhell-adsds}}
\end{figure}

Most existing games and simulations, including non-Euclidean ones, do not take relativity into account. At every point, they represent the current state of the simulation (at time $t$)
in the computer memory for all objects using the chosen internal model of the geometry, and to compute the further states of the simulation (at times $t'>t$), 
a chosen model of physics is used, usually some simplification and adaptation of Newtonian physics. 
Interestingly, while such a non-relativistic model of Euclidean spacetime obeys the Galilean principle of relativity (a moving object having no access to external
reference point cannot determine that it is moving), this is no longer the case for non-Euclidean geometries. For example, in $\bbH^d$, 
lines diverge, so a large object moving experiences apparent centrifugal force. This is the reason why, for example, 
the flock of boids in the flocking simulations in RogueViz \cite{rogueviz2023} cannot keep its shape, contrary to a Euclidean flocking simulation \cite{boids}.
For a similar reason, HyperRogue \cite{hyperrogue} could not feature large, freely moving objects -- the boundary of such an object would have to move significantly faster than the center, 
in a curved line, which would be unintuitive; so while the game does feature some somewhat large creatures (such as snakes and krakens), they are narrow enough to avoid this problem.

This can be solved by using \emph{de Sitter} ($\dS{d}$) and \emph{anti-de Sitter} ($\uadS{d}$) spacetimes,
which are relativistic analogs of $\bbS^d$ and $\bbH^d$, respectively. They are symmetric under the Lorentz transformations
used to simulate the change of velocity, and therefore, they obey the Galilean principle of relativity. Intuitively, while
$\bbH^d$ and $\bbS^d$ stretch the distances in space, $\dS{d}$ and $\uadS{d}$ also stretch
the time. The game \emph{Relative Hell} (Figure \ref{relhell-adsds}) lets the player fly a two-dimensional spaceship in these spacetimes.
Due to the nature of these spacetimes, the gameplay is different: in $\dS{2}$, objects are naturally
pulled apart, so the goal is to keep close to the \emph{main star} while avoiding bullets, as in the
\emph{bullet hell} genre; while in $\uadS{2}$, objects are naturally pulled together, so the whole space 
is rotating in order to generate centrifugal force to balance that, and the goal is to shoot, similar to
classic omnidirectional shooters such as \emph{Asteroids}. The player can also gain insight into
the usual non-Euclidean geometric and relativistic (time dilation, space contraction) phenomena, such 
as the exponential expansion of $\bbH^2$, and time dilation and Lorentz contraction.

\section{Preliminaries}\label{ref:prelim}
We briefly recall the definitions of basic spaces and spacetimes. For more details, 
see, e.g., \cite{cannon} for hyperbolic geometry,
and \cite{GriffithsPodolsky}
for $\dS{d}$ and $\uadS{d}$.

The \emph{Euclidean space} of dimension $d$, $\bbE^d$, is $\bbR^d$ equipped with the Euclidean inner
product $g_E(x, y) = \sum_i x_i y_i$. For a vector $x \in \bbR^d$,
we use the notation $x[a/i]$ for the vector $x$ with $i$-th coordinate replaced by $a$. In particular, $0[1/i]$
is the point whose $i$-th coordinate is 1 and the other coordinates are 0.
Our Euclidean inner product defines the distance between points: $d_E(x, y) = \sqrt{g_E(x-y, x-y)}$.
The length of a differentiable curve $\gamma: [a,b] \ra \bbE^d$ can be computed as 
\(
\int_a^b \sqrt{g_E(\dot \gamma(t), \dot \gamma(t))} dt\label{curveeq};
\)
note that $d_E(x,y)$ is also the length of the shortest curve from $x$ to $y$.
Orientation-preserving isometries of the Euclidean space are generated by translations $T_i^a(x) = x[x_i+a/i]$ and 
rotations $R_{i,j}^\alpha(x) = x[x_i\cos \alpha + x_j\sin \alpha/i][x_j\cos\alpha-x_i\sin\alpha/j]$ for $i \neq j$.
Isometries preserve $g_E$, and thus distances and curve lengths.
The basic translations and rotations can be composed to obtain other isometries. Isometries of $\bbE^d$
correspond to looking at $\bbE^d$ from another frame of reference.

In video games and computer graphics, it is convenient to use the \emph{homogeneous coordinates}, that is,
$\bbR^{d+1}$, where the extra $(d+1)$-th coordinate is always equal to 1. This method lets us represent 
translations and rotations as matrices.

The \emph{spherical space (sphere)} of dimension $d$, $\bbS^d$, is $\bbE^{d+1}$ restricted to $\{x: g_E(x,x) = 1\}$. (For convenience
we consider only spheres of radius 1 here.)
The distance $d_S(x,y)$ between two points $x,y \in \bbS^d$ is the length of the shortest curve (geodesic)
connecting them on the sphere, $c: [a,b] \ra \bbS^d$; we have $d_S(x,y) = \acos(g_E(x,y))$.
Orientation-preserving isometries of the sphere are generated by $R_{i,j}^\alpha$ (while in the Euclidean space we needed the extra
coordinate for translations, in $\bbS^d$ we can instead just use the one extra coordinate we already have). A sphere is maximally symmetric: any isometry of the underlying $\bbE^{d+1}$
that maps 0 to 0 maps the sphere to itself.

A \emph{signature} is a $\sigma \in \{-1,+1\}^d$; we will drop the 1 and just write the signs, for example $(+,+,+,-)$; if a sign
repeats, we use an exponent, for example $(+^3,-)$.
The \emph{pseudo-Euclidean spacetime} of signature $\sigma$, $\bbE^\sigma$, is $\bbR^d$ equipped with the inner product
$g_\sigma(x,y) = \sum_i \sigma_i x_i y_i$. In this paper, coordinates $x_i$ with $\sigma_i=1$ correspond to space dimensions,
and coordinates with $\sigma_i=-1$ correspond to time dimensions. Positive values of $g_\sigma(v,v)$ correspond to
space-like intervals, and negative values correspond to time-like intervals. A curve is \emph{space-like} if the integral in equation (\ref{curveeq}) is
well-defined (that is, the value under the square root is always non-negative), and \emph{time-like} if this value is always non-positive. The proper time of a time-like curve $\gamma$ is defined as $\int_a^b \sqrt{-g_E(\dot \gamma (t), \dot \gamma (t))} dt$.

In special relativity, our spacetime is modelled as a pseudo-Euclidean
space of signature $(+,+,+,-)$; the fourth coordinate corresponds to time, and the units of time and distance are
chosen so that Einstein's constant $c$ equals 1. Isometries of pseudo-Euclidean spacetime are similar, but for $\sigma(i) \neq \sigma(j)$
we replace rotations $R_{i,j}^\alpha$ by Lorentz boosts $L_{i,j}^\alpha(x) = x[x_i\cosh \alpha + x_j\sinh \alpha/i][x_j\cosh\alpha+x_i\sinh\alpha/j]$.
The parameter $\alpha$ is called \emph{rapidity}; in spacetimes with 1 time coordinate, Lorentz boosts correspond to changing the velocity of our
frame of reference. 
Objects generally move along time-like curves; Proper time is interpreted as the time measured by a clock, or intuitively felt by a sentient creature, moving along such a curve.
Time-like geodesics have the longest possible proper time. While the relative speed of moving objects $\tanh\alpha$ is bounded by $c=1$, the rapidity is not bounded.
The interaction of space and time coordinates causes well-known relativistic effects such as Lorentz contraction and dilation of time.
The set of points connected to $v$ with time-like geodesics $\gamma$ such that the time coordinate of $\gamma(t)$ is increasing on is called the \emph{future light cone} of $v$, and
the set of points connected with time-descreasing time-like geodesics is the \emph{past light cone}. These light cones are preserved when we apply
the isometries of $\bbE^\sigma$ that keep $v$. The other points are neither in the future or the past, but rather elsewhere -- their time coordinate
may be greater or smaller than the time coordinate of $v$, depending on the chosen isometry. What happens at $v$ causally depends only on the spacetime events
in the past light cone of $v$, and may affect only the spacetime events in the future light cone of $v$.

The \emph{hyperbolic space} of dimension $d$, $\bbH^d$, is $\bbE^\sigma$ for $\sigma = (+^d, -)$ restricted to $\{x: g_\sigma(x,x)=-1, x_{d+1}>0\}$.
This is the Minkowski hyperboloid model of hyperbolic geometry. This is a space, not a spacetime, that is, all the curves on $\bbH^d$ are
space-like. While hyperbolic geometry is usually taught in the Poincar\'e disk model, the Minkowski hyperboloid model is generally easier to
understand (assuming familiarity with Minkowski geometry) and work with computationally because of its similarity to the natural model of the sphere. In particular, orientation-preserving
isometries are generated by rotations $R_{i,j}^\alpha$ and Lorentz boosts $L_{i,d+1}^\alpha$ (corresponding to translations of $\bbH^d$),
and the distance between $x$ and $y$ is $\acosh(g_\sigma(x,y))$. While in special relativity the time coordinate is usually indexed as $x_0$,
in this paper we prefer to make it the last coordinate, for consistency with the usual indexing of homogeneous coordinates of the Euclidean space in computer graphics.

Taking $z \in \bbR$, we can project $x\in \bbH^d$ to $x' \in \bbR^d$ by $x'_i = x_i/(x_{d+1}+z)$. This is called the \emph{general perspective projection}.
For $z=1$ we get the \emph{Poincar\'e ball model} (also called the \emph{Poincar\'e disk model} in 2 dimensions).
It is the hyperbolic analog of the stereographic projection of the sphere, which uses the same formula. It is the
most popular method of visualizing hyperbolic geometry; just like the stereographic projection, it is conformal, meaning that the angles and small shapes are mapped faithfully. 
Another one is the \emph{Beltrami-Klein ball (disk) model}, obtained for $z=0$. Beltrami-Klein disk and Poincar\'e disk are called models because they
are alternative mathematical representations of hyperbolic geometry; in this paper, we use only the Minkowski hyperboloid model for mathematical representation, and the disk models
are used as projections for visual representation.
More possible projections exist which are not special cases of the general perspective projection, for example the azimuthal equidistant projection, which renders the distances and angles
from the chosen central point correctly \cite{visiccs}. Dozens of spherical projections
are used in cartography \cite{snyder1997flattening}; many of them have hyperbolic analogs, available in the HyperRogue engine \cite{hyperrogue}.

The \emph{de Sitter spacetime} with $d$ space dimensions and 1 time dimension, $\dS{d}$, is $\bbE^\sigma$ for $\sigma = (+^{d+1}, -)$ restricted to 
$\{x: g_\sigma(x,x)=1\}$.

The \emph{wrapped anti-de Sitter spacetime} with $d$ space dimensions and 1 time dimension, $\wadS{d}$, is $\bbE^\sigma$ for $\sigma = (+^d, -^2)$ restricted
to $\{x: g_\sigma(x,x)=-1\}$. Note that $ad\bbS^d$ has closed time-like loops, for example, $\gamma(t) = 0[\sin t/{d+1}][\cos t/{d+2}]$. Such closed-time
loops are obtained by going around the axis $A = \{x \in \bbE^\sigma: \forall i\leq d\ x_i=0 \}$. Due to the closed time-like loops, $\wadS{d}$ has no causal structure
(everything is simultaneously in the past and future of everything else and itself). This problem is resolved by taking the universal cover of
$\wadS{d}$, i.e., the point reached by going $n\neq 0$ times around the axis $A$ is considered a different point in the spacetime. We will call this universal cover
the \emph{unwrapped anti-de Sitter spacetime} $\uadS{d}$, or just \emph{anti-de Sitter spacetime} for short.

Spacetimes $\dS{d}$ and $\uadS{d}$ will be discussed in detail in the following sections. Here, we will only remark that time-like and space-like curves
and their lengths and proper times are defined as above, and that both $\dS{d}$ and $\uadS{d}$ are maximally symmetric spacetimes, with their isometries 
generated from $R_{i,j}^\alpha$ %
and $L_{i,j}^\alpha$
.

\section{Simulation of the Anti-de Sitter spacetime}
For simplicity, we will start with $\wadS{d}$. When necessary, we fix $d=2$.

The point $O = 0[1/d+2]$ is considered the origin of the $\wadS{d}$. Every object $b$ in the simulation, at a specific point of its proper time $t$, is represented by an isometry
$T_{b,t}$ of $\wadS{d}$, which maps the coordinates relative to $(b,t)$ to the world coordinates. Usually, $(b,t)$-relative coordinates of $b$ itself at time
$t$ is $O$, so the world coordinates of $b$ at time $t$ are $T_b(O)$. The player controls a ship $s$, which is one of the objects in the game. To display an object
at world coordinates $x$ on the screen at time $t$, we need to map $x$ into ship-relative coordinates, $x' = T_{s,t}^{-1} x$. The state of the
game universe at the current time is a slice of the spacetime. This slice is $S = \{x \in \wadS{d}: x_{d+1} = 0, x_{d+2}>0\}$, which is isometric to $\bbH^d$, and can be rendered e.g.~in the
Poincar\'e disk/ball model.

The ship $s$ is controlled in real time by a player, so for every time moment $t$ displayed in an animation frame, the game has to compute
the next frame transform $T_{s,t+\epsilon}$ depending on both $T_{s,t}$ and player's decisions. Objects in spacetime move along timelike geodesics if no force is acting on them. Timelike geodesics
in $\wadS{d}$ are generally of form $T R^t_{d+1,d+2} O$. Thus, if the player does not accelerate from time $t_1$ to $t_2$, we have $T_{s,t_2} = T_{s,t_1} R^{t_2-t_1}_{d+1,d+2}$.
Changing the camera speed in dimension $i\in\{1,\ldots,d\}$ corresponds to changing the frame of reference by multiplying it by $L^\alpha_{i,d+2}$. Thus, if the player accelerates, we also need to
multiply the formula for $T_{s,t_2}$ by $L^\alpha_{i,d+2}$ on the right.

One possible objection to displaying the slice $S$ is that the player should not see the state of the universe at the current time -- the ship at time $t$ only knows the past light cone of
$T_{s,t}O$. In the current game prototype, we assume that all the objects other than $s$ behave deterministically, so
this is not an issue---the ship could compute the current state of other objects depending on what it knows. While displaying $S$ is less immersive, it is useful for understanding how
the spacetime works. (One can change the options to get the actual view.)

The simplest deterministic movement is geodesic movement. Let $b \neq s$. The object $b$ at any given time is not a point, but it is rather a subset $X(b) \subseteq S$. 
The formula $T_{b,t} = T_{s,0} R^t_{d+1,d+2}$ turns out not to be
satisfactory -- while the object $b$'s origin $O$ moves geodesically, other points in $X(b)$ do not. In $d=2$, this issue can be fixed by using the formula $T_{b,t} = T_{b,0} R^{t}_{3,4} R^{t}_{1,2}$,
which enforces geodesic movement of every point of $b$. For short, we denote the isometry $R^{t}_{3,4} R^{t}_{1,2}$ with $M^t$. (It is possible to change the options to also use $M^t$ for the ship,
but with such a setting, the game becomes more confusing to the player.)
Every object $b \neq s$ starts its lifetime at some time $t_1$ and ends at time $t_2$; the time $t_1$ may be $-\infty$ if the object did exist forever,
and $t_2$ may depend on player's actions, and be $\infty$ if it is not yet known to be ever destroyed. We can compute $T_{b,t}$ at every time $t$ knowing $T_{b,0}$.
So, to display the point $x \in X(b)$ of object $b$ at the ship's proper time $t$, we need to find $t_b$ such that $x_b := T^{-1}_{s,t} T_{b,0} M^{t_b} x \in S$. If $t_b \in [t_1, t_2]$, we apply the chosen
projection to $x_b$. We display the point $x$ there.

So far, we have been assuming $\wadS{d}$ for simplicity, which does not really work due to the time-like loops -- technically, all of spacetime is in the past, so the player would be able to perceive
the results of the actions they have not performed yet (this is a problem with all games featuring a powerful enough form of player control and time travel). Another issue is numerical precision:
hyperbolic space ($S$ in our case) is characterized by its exponential growth, which causes numerical errors to accumulate quickly, and using world coordinates relative to some fixed origin does not
really work when we travel far enough from that origin. Our solution of these issues is based on the existing implementation of $\widetilde{G}$, the universal cover of Lie group $G$ of orientation-preserving isometries
of $\bbH^2$, also called  $\widetilde{SL(2,\bbR)}$ or the twisted product of $\bbH^2$ and $\bbR$. This implementation is described in paper \cite{rtvizfinal}. The space of isometries of $\bbH^2$ is
a three-dimensional space; the three dimensions correspond to the two dimensions of $\bbH^2$ itself (translations), plus one extra dimension which corresponds to rotation by some angle $\alpha$. 
In \cite{rtvizfinal}, this space of rotations is represented using unit split quaternions (this is the hyperbolic analog of the fact that isometries of $\bbS^2$ are represented
using quaternions, which is well known in computer graphics). The set of unit split quaternions corresponds exactly to $\wadS{d}$ -- the only difference is that in \cite{rtvizfinal} the dimension corresponding
to rotation is considered spacelike, while in $\wadS{d}$ it is timelike; specifically, rotation by angle $\alpha$ corresponds to $M^\alpha$\longonly{ defined in the last paragraph}.

In the universal cover, we consider the isometries whose rotation components are described by different angles $\alpha, \beta \in \bbR$ to be different, even if they are actually the same rotation
(that is, $\alpha-\beta$ is a multiple of $2\pi$). We have a natural projection of the universal cover to the underlying space $\pi: \uadS{d} \ra \wadS{d}$. The space $\uadS{d}$ has its origin $O'$, 
$\pi(O') = O$. We pick a lift $j: \wadS{d} \ra \uadS{d}$ such that $\pi \circ j$ is identity; the point $j(x)$ is chosen among the points $x'$ such that $\pi(x')=x$ in a natural way: 
the path from $x'$ to $O'$ does not cross $\pi^{-1}\{x \in \wadS(d): x_1<0, x_2=0\}$. We reuse the same notation $M^\alpha$, $R^\alpha_{i,j}$ and $L^\alpha_{i,j}$ for the isometries of $\uadS{d}$.
Note that the pass-of-time isometry $M^\alpha$ (as well as the underlying $R^\alpha_{2,3}$) will now be different for every $\alpha \in \bbR$, while in $\wadS{d}$ we had $M^{2\pi} = M^0 = Id$.
If $T$ is an isometry of $\uadS{d}$, we likewise have uniquely defined $\pi(T)$, which is an isometry of $\wadS{d}$ such that $\pi(Tx) = \pi(T)(\pi(x))$. If $T$ is an isometry of $\wadS{d}$, let
$j(T)$ be the isometry of $\uadS{d}$ which takes $O$ to $j(T(O))$ and such that $\pi(j(T)) = T$.

\def\dsh{h}

In our simulation engine, the points of $\widetilde{G}$, and equivalently $\uadS{d}$ are represented as \emph{shift points}. A shift point $(x,\dsh)$ consists of a point $x \in \wadS{d}$ and a shift $\dsh \in \bbR$, and represents
$x' = M^{\dsh} j(x)$. Note that this representation is not unique; the \emph{canonical} representation of $x'$ is the one in which $x_3=0$ and $x_4>0$. Similarly, a isometry $T'$ of $\uadS{d}$ is
represented as \emph{shift matrices}: $(T,\dsh)$, where T is an isometry matrix and $\dsh$ is a shift, represents $M^{\dsh} j(T)$. The canonical representation of $T'$ is the one in which $(TO, {\dsh})$ is 
canonical.

To avoid the numerical precision issues, we tessellate $\uadS{d}$. That is, $\uadS{d}$ is subdivided into a number of tiles $\tau$, and every object is described not by a shift matrix relative
to some origin of the whole space, but by a tile $\tau$ it lives in, and a shift matrix $(x,d)$ relative to the center of the tile $\tau$. As previously mentioned, $\uadS{2}$ corresponds to
the Lie group of isometries of $\bbH^2$; the centers of our tiles correspond to the isometries which map the order-3 heptagonal tessellation (Figure \ref{fig73}) to itself. Knowing the shift matrices transforming
the coordinates relative to tile $\tau_1$ into the coordinates relative to adjacent tile $\tau_2$, it is straightforward to compute the coordinates of all the object nearby to the player, 
relative to the player. The tessellations themselves can be computed using discrete methods such as automata theory \cite{wpigroups,gentes}, thus avoiding numerical precision issues arising
otherwise when the player makes their ship travel a large distance from the start.

Due to the nature of time-like geodesics in $\uadS{d}$, objects appear to move in circles. This suggests a game design somewhat reminiscent of the classic arcade multidirectional shooter game
{\it Asteroids} (which took place in a space with torus topology, so the objects also did not escape the playing area). The player has to shoot rocks for resources, such as gold (increasing score), health (used up
when hit by an asteroid), ammo (used up when shot), fuel (used up when accelerating), and oxygen (used up proportionally to proper time elapsed). These are standard resources well-known to
gamers. One interesting consequence of relativity is that the player may use up the fuel resource to save a bit of the oxygen resource, since acceleration can be used to reduce the proper
time necessary to reach another point in the spacetime (similar to the twin paradox). Shooting the missile creates a new object $m$ (missile); we compute if the worldline of the missile $m$ intersects the world line
of some rock $r$, and if so, the life of both $m$ and $r$ end at this time, and we create a collectible resource at the same spacetime event.

The implementation of such a simulation requires us to implement the necessary operations for shift points and shift matrices: compose isometries (multiply shift matrices), apply isometry to a point (multiply shift matrix by a 
shift point), find $t_b$ and $x_b$ for rendering objects. While in $\wadS{d}$ we would just use the well-known matrix and vector multiplications, in $\uadS{d}$ these operations are somewhat more involved 
due to the necessity of computing shifts correctly (we need to be careful to obtain the correct shift value $h$ instead of, e.g., $h \pm 2\pi$). To keep the paper short, we do not include the full formulas in this paper; they can be found in the source code of our simulation
(file {\tt{math.cpp}}).

\section{Simulation of the de Sitter spacetime}
The general ideas of our de Sitter simulation are similar to the anti-de Sitter case described in the previous section, so we only list the differences.

The origin is now $O = 0[1/d+1]$. The slice corresponding to current time is $S = \{x \in \dS{d}: x_{d+2}=0\}$. It is isometric to $\bbS^d$, so it can be rendered e.g.~in the stereographic model.
The pass of time is now represented using isometry $M^\alpha = L^\alpha_{d+1,d+2}$ (we use the same isometry for the pass of time for $s$ and the other objects). If the player accelerates,
we multiply $T_{s,t_2}$ by $L^\alpha_{i,d+2}$.

The equivalence of $S$ to $\bbS^d$ might suggest {\it Asteroids}-like design again: rocks flying around the sphere. However, the spacetime $\dS{d}$ works very differently: the space appears to be
expanding as time passes, and objects which fly too far away from $s$ are impossible to reach anymore. In particular, the other side of the sphere is unreachable. If the game started with a number
of randomly pre-placed objects at time 0, after some time all of them would depressingly fly away from each other, with at most one of them remaining in the part of universe reachable to the player.
So the world of our de Sitter game is constructed differently. There is a main star (black-and-yellow in Figure \ref{relhell-adsds}), and the goal is to remain close to the main star as long as possible.
Other objects in the universe make this task harder. Every object (bullet) $b$ gets close to the main star at some time $t_b$; avoiding being hit is the main challenge of the game,
thus making the game an example of a game in the \emph{bullet hell} genre. As usual in bullet hell games, the bullets arrive in waves (sharing similar value of $t_b$) arranged in various patterns.

Numerical precision issues caused by the space expanding exponentially also arise in $\dS{d}$. This time, the solution we use in our simulation is not generally applicable, but rather tailored to the design of our game,
that is, based on the assumption that the player will be always forced to remain close to the main star. We again use the concept of shift matrices: the shift matrix describing an object $b$ will be of form
$(T, t_b)$, which is equivalent to matrix $M^{t_b} T$ relative to the main star at time 0, or equivalently, $T$ relative to the main star at time $t_b$. Objects need not be rendered if their
$t_b$ is not close enough to the proper time of the main star currently visible to the player (if we tried to render them, we would run into precision issues due to the exponential
growth of $\sinh$ and $\cosh$).

In the anti-de Sitter case we assumed that every point of the object $b$, $x \in X(b) \subseteq S$, moves geodesically. This does not work in $\dS{d}$ -- if every point moved geodesically, the objects would
expand. Instead, we only assume that the center $O$ of the object moves geodesically. Instead, after computing $t^O_b$ and $x^O_b$ for the center $O$ ($x_b = T^{-1}_{s,t} T_{b,0} M^{t_b} O \in S$), for every other
point $x \in X(b)$ we find a geodesic $\gamma'_x$ which is correct at time $t_b$ using the formula $\gamma'_x(t_b+t) = T_{b,0} M_{t_b} L^{x_1}_{1,4} L^{x_2}_{2,4}$ where $(x_1,x_2)$ are the coordinates of point $x$, 
and find $x'_b$ and $t'_b$ such that $x'_b = T^{-1} {s,t} \gamma'_x(t'_b)$. The point is now rendered at $x'_b$. We compute $x'_b$ for every vertex of the polygonal model describing the object $b$, and render $b$
as a polygon with vertices obtained by mapping $x'_b$ using the chosen projection.

\section{Visualizations and Insights}

\begin{figure}[t]
\begin{center}
\includegraphics[width=.32\linewidth]{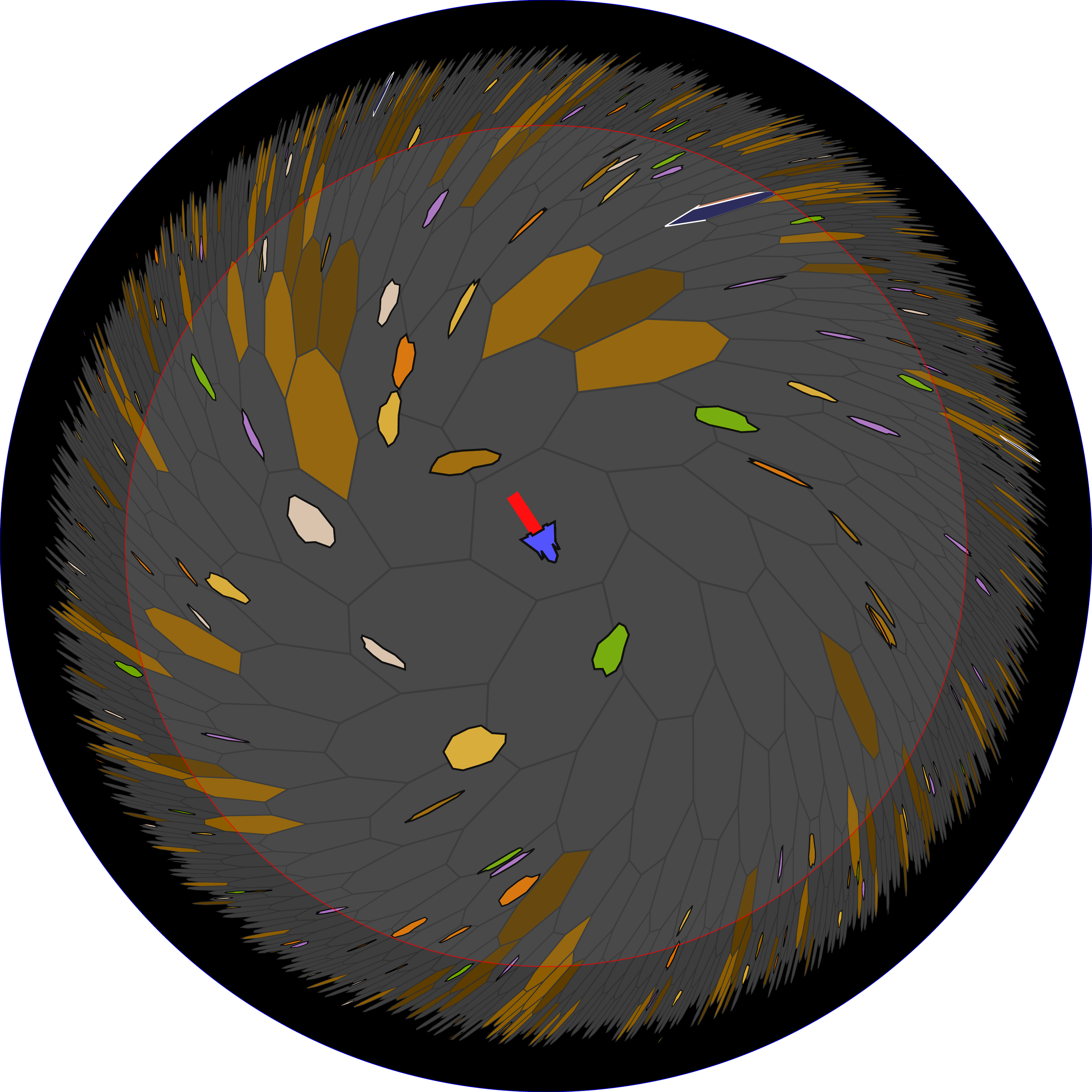} \includegraphics[width=.32\linewidth]{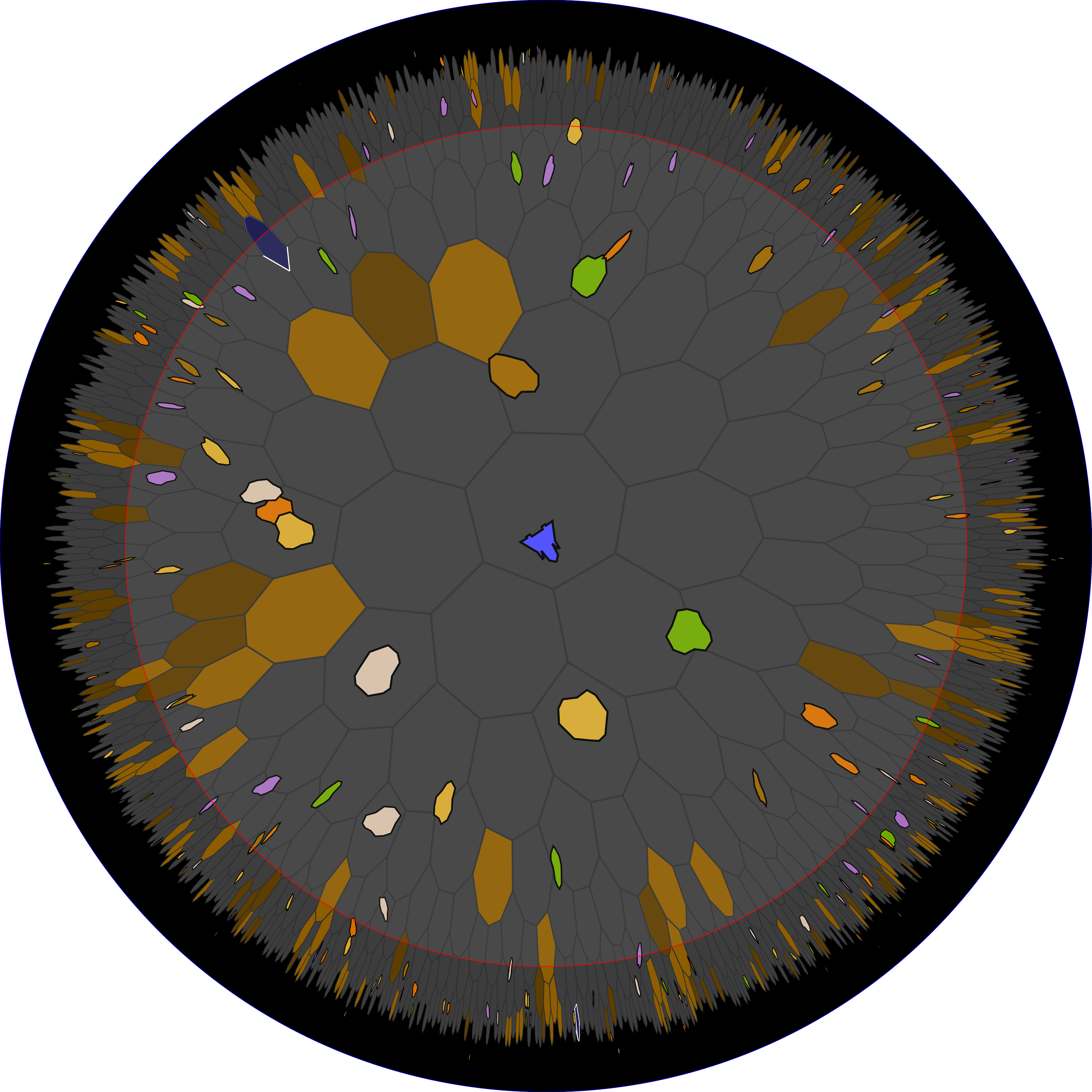} \includegraphics[width=.32\linewidth]{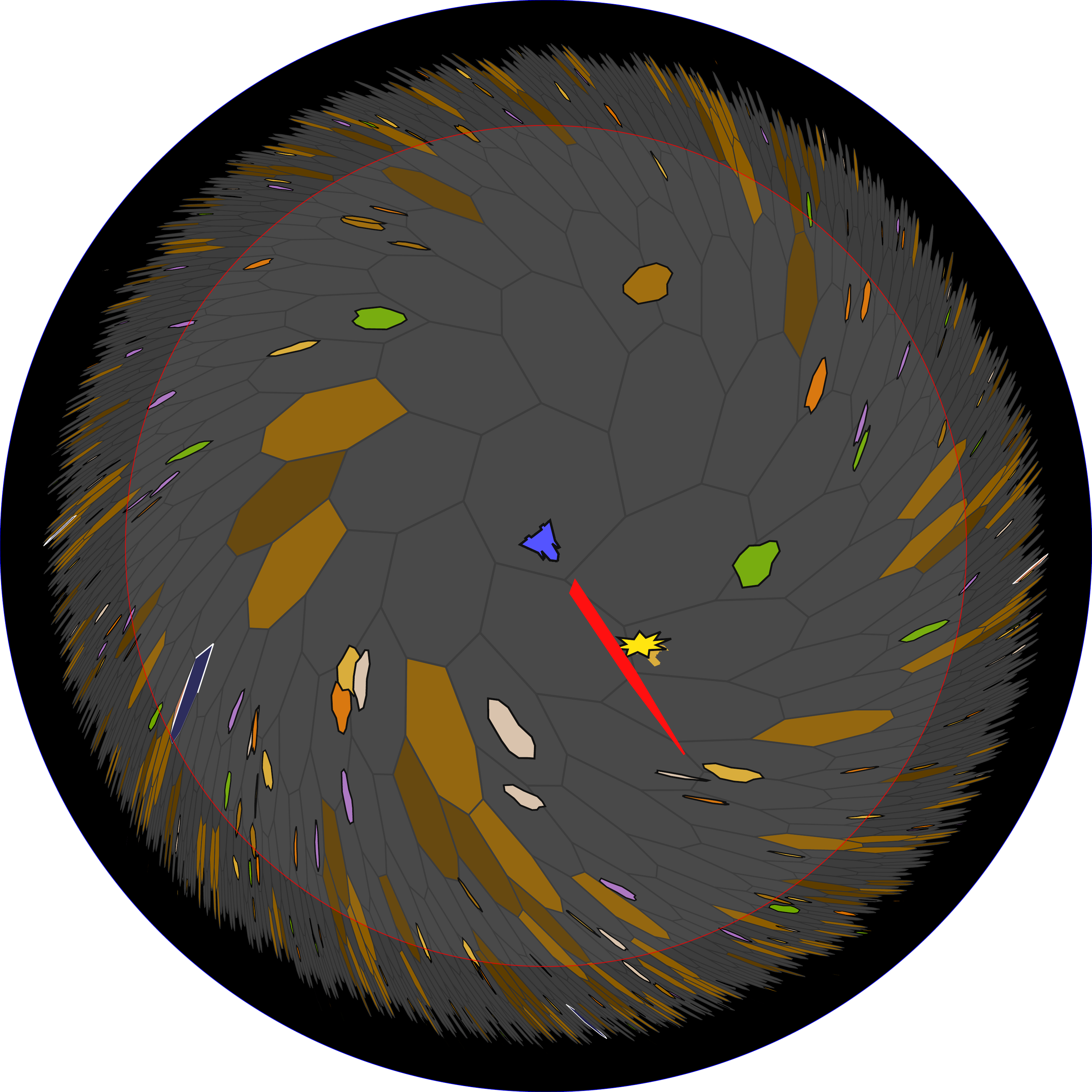}
\end{center}
\caption{Anti-de Sitter spacetime: past light cone view (left), present (middle), and future light cone view (right). Note the stretched missile in the future light cone view.
Taken from a replay; the red circle is the boundary of the light cone relative to a future position of the ship. Poincar\'e disk model.
\label{coneview}}
\end{figure}

\begin{figure}[t]
\begin{center}
\includegraphics[width=.32\linewidth]{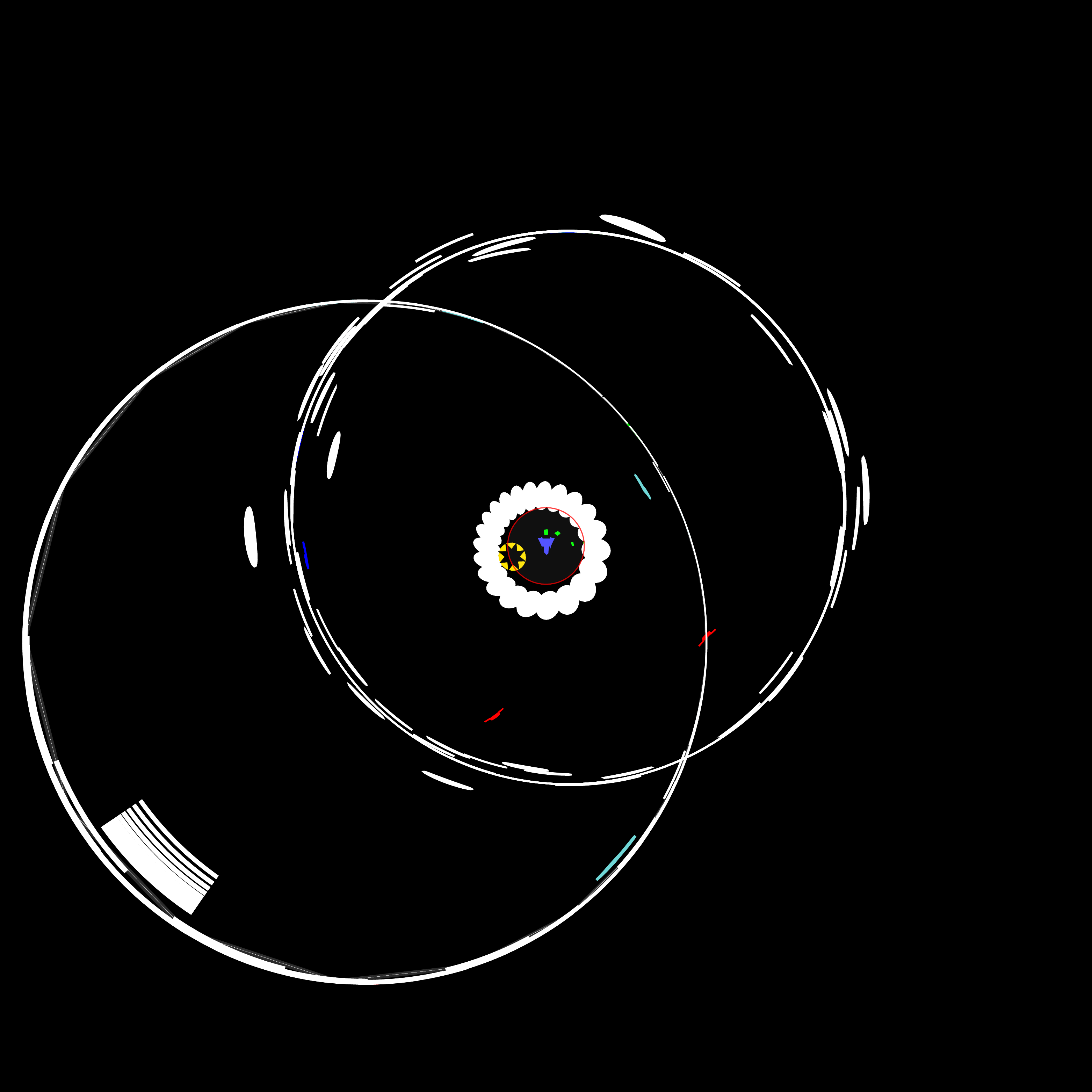} \includegraphics[width=.32\linewidth]{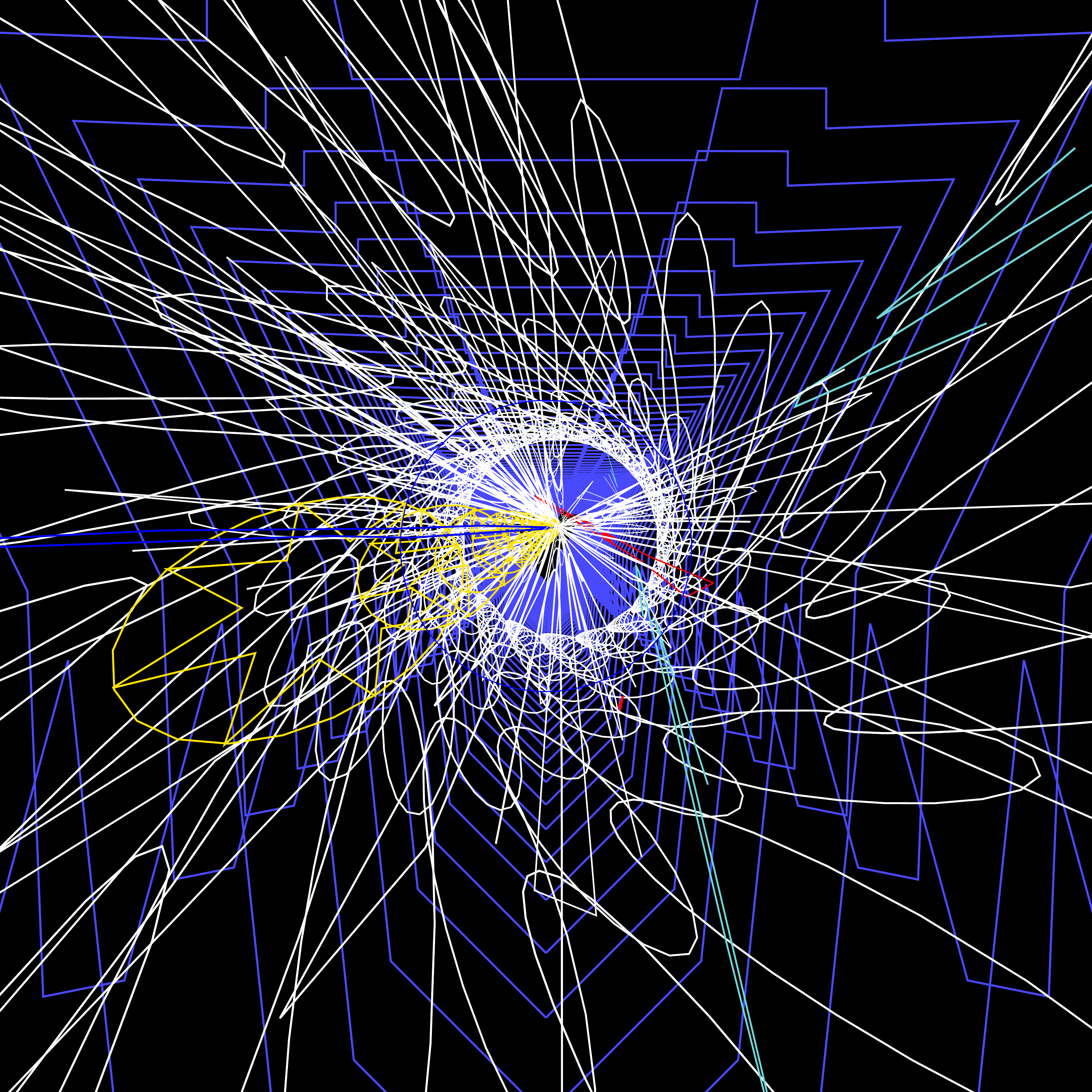} \includegraphics[width=.33\linewidth]{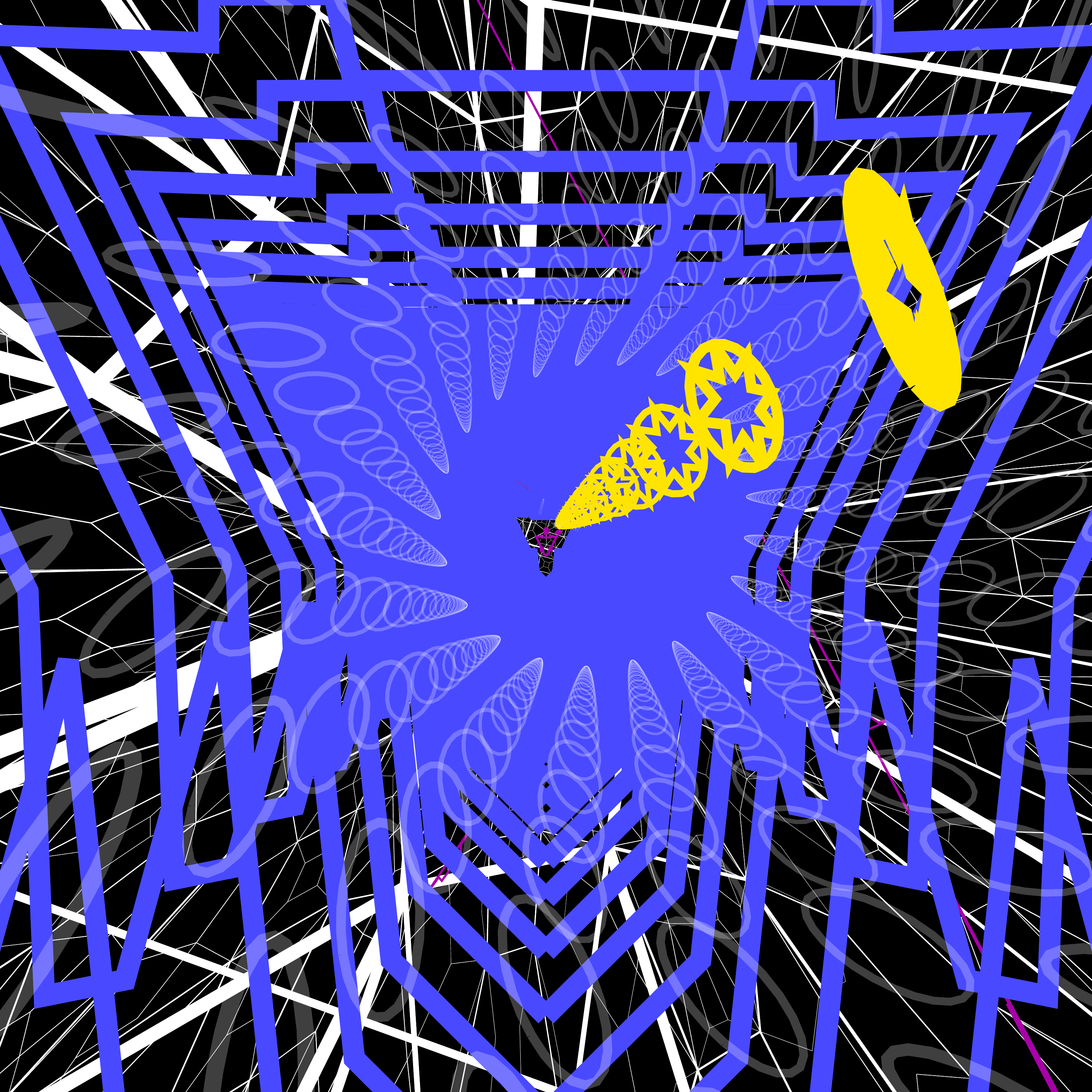}
\end{center}
\caption{De Sitter spacetime: present in stereographic projection (left), spacetime view (center), and $\bbH^3$ view (right). Unfortunately, the spacetime and $\bbH^3$ views are difficult to capture
in a still image (and also in video due to the compression).\label{spacetimeview}}
\end{figure}

Science-based games should allow the player to not only experience the challenge provided by the gameplay, but also play with the simulation. One aspect of this is that 
the player can change the parameters of the game (amount of resources, rocks, scale, speed, etc.) to explore more diverse scenarios without being bothered with the challenge,
making the game accessible to players who are not skilled at the given game genre.
Another aspect is that we should keep the history of every object in the game. Such history keeping is necessary by the basic construction of a relativistic game -- for example,
even if the game knows that some object $b$ disappears at proper time $t$ (for example, a rock is hit by player's missile), and the player has seen that, the player might accelerate,
causing them to see the object $b$ still existing, due to how Lorentz transformations work. While it would be safe to remove $b$ from memory if the life of $b$ ended in the past light cone,
for the discovery purposes it is better to just remember everything. Since the whole history is kept, the player can replay their game,
possibly from other frame of reference. 
While viewing such a replay (or possibly even during the actual game), the player can change
aspects of how the in-game universe is visualized, such as:

\begin{itemize}
\item In $\uadS{d}$, let the time $\alpha$ pass according to $R^{\alpha}_{d+1,d+2}$ or $M^{\alpha}$. The first option is generally more easier to understand (the camera is not rotating),
but both views are interesting. In the second view, the world appears to stop spinning around the player, but the movement starts looking like wrapping the spacetime, which is interesting but unintuitive.

\item The player can choose to use the current $S$ as explained so far, or they can take $S$ to be the boundary of the past light cone (corresponding to the events the ship would be actually seeing
at that precise moment), or the boundary of the future light cone (Figure \ref{coneview}).

\item Proper time of every object could be displayed, to let the player see whether the relativistic effects (time dilation) work as they expect.

\item There is also an option to view the 2+1-dimensional spacetime using the perspective projection. In this visualization, it is assumed that light travels along geodesics of all kinds. We see
a circle of radius 1, the interior of that circle corresponds to time-like geodesics (space-time events the ship could reach), and the exterior corresponds to events happening elsewhere. The circle
is essentially a hyperbolic plane in the Beltrami-Klein model; changing the ship's velocity transforms the interior of the circle according to the isometries of this hyperbolic plane.
See Figure \ref{spacetimeview} for an example.

\item Interestingly, the isometries of $\dS{d}$ are also the isometries of $\bbH^3$. Both of them live in $\bbE^\sigma$ for $\sigma=(+^3,-)$, but $\bbH^3$ is the set of points $x$ where
$g_\sigma(x,x)=-1$, while $\dS{2}$ is the set of points $x$ where $g_\sigma(x,x)=1$. This observation leads to another visualization: take the isometry describing the ship's view, 
and use the same isometry to view $\bbH^3$ using Klein-Beltrami ball model, or equivalently, perspective. Such a dual view has interesting properties (pass of time corresponds to moving the
dual camera forward; accelerating corresponds to moving the dual camera sidewise; points of $\dS{d}$ are represented in this perspective as points outside of the Klein-Beltrami ball, in particular
points of $S$ being represented as points in infinity; etc.). See Figure \ref{spacetimeview} for an example.

\end{itemize}

The following insights into de Sitter and anti-de Sitter spaces are gained.

\begin{itemize}
\item The space in the anti-de Sitter game appears to be rotating. This is because the fixed parts of the map (the walls based on the tessellation) move along the geodesics $M^{t_1} v$, while the 
progress of time is modelled by multiplying the spacetime coordinates $T^t_{3,4}$. So, if $v \in S$ and the player moves by $t$ units, the obbject is displayed at (taking $t=t_1$) $T^{-t}_{3,4} M^t v = 
R^{1,2}(t)$. An object cannot simply stay in a fixed place $v$ on screen, because that would not be geodesic movement: the time interval between $v \in S$ and $T^{t}_{3,4}v$ is greater and greater
when we increase the distance from $v$ to $O$, and thus timelike geodesics are pulled towards the center. So in general objects appear to be orbiting around the ship in circular or elliptical orbits.

\item In the de Sitter game, if we imagine $S$ as a sphere with the main star at the north pole $N$, objects moving along geodesics will get close to $N$ at some point, then escape towards the equator.
Similarly, if we reversed the time, we discover they started towards the equator too. If we view the spacetime relative to the main star, the equator is well visible due to all the objects close to it.
Normally we view the spacetime relative to the ship, which is a different frame of reference, so the two equators are distinct. Figure \ref{relhell-adsds} shows one of them, while Figure \ref{spacetimeview}
shows the equator circle made by bullets we have avoided in the past, the equator circle made by bullets we will have to avoid in the future, and also a small circle of bullets we are avoiding right
at the moment when the screenshot was taken.

\item Another observation is the Lorentz contraction. All objects appear compressed in the direction of the movement. For example, while all the tessellation tiles are of regular heptagonal shape
and the Poincar\'e model is conformal, but they appear more and more narrow as we move further away from the center (Figure \ref{relhell-adsds}). Due to the relationship between split quaternions and
the underlying hyperbolic plane, the distance between tiles in $\uadS{2}$ are half the distances between the respective tiles in $\bbH^2$; after switching the view to Beltrami-Klein model, the 
heptagons look regular again.

\item Dilation of time is best observed in the de Sitter game, by comparing the proper time of the ship with the score, which is the proper time of the main star we are close to. Generally the 
score will be smaller, due to the geodesic movement of the main star, and non-geodesic movement of the ship. Interestingly, when the player loses the de Sitter game
by getting too far away from the main star, their score no longer increases -- the main star is escaping so fast that its proper time seen by us does not change.
\end{itemize}

\section{Further work}
The engine described in this paper can be extended to other games and experiments in $\uadS{2}$ and $\dS{2}$. To conclude the paper, we describe some extensions.

\begin{itemize}
\item An active enemy that attempts to predict where the player's ship is going to go (for example, assuming that the ship is simply going along a time-like geodesic), and shoot there. Such enemies are popular in video games.
This concept would be naturally more interesting in a relativistic game, since such an enemy would have to predict the player's ship position based on the past (that is, if the enemy shoots at time $t$, it only sees
what the player did at some time in the past, due to the limited speed of light). This is still deterministic. Non-deterministic enemies are of course also a possibility. However, interestingly, making the game multi-player
introduces an unexpected challenge, at least if we want the proper time of both players proportional to the real time elapsed. One interesting way to avoid this issue would be to make the game end in the case when causality
is lost, i.e., one of the players falls inside the past light cone of another player.

\item A map editor to let the players construct their own scenarios and puzzles, for example, to explore the twin paradox (using fuel to conserve oxygen).

\item We have chosen the game to have two spatial dimensions. Two-dimensional games generally showcase the experimental gameplay better (multi-directional shooters and bullet hell games work better in two dimensions),
and also allow our three-dimensional visualizations of the space-time. However, immersive three-dimensional variants of our games would also be worthwhile. Our implementation of $\uadS{2}$ crucially depends on two-dimensionality,
in particular, the trick of using $M^\alpha$ to pass time while keeping all the world stable, due to all dimensions being rotated, works only if the number of space dimensions is even. In three dimensions, one of
the dimensions would not be rotated, and thus not stable. A simple way to add another dimension, keeping our tessellations of $\bbH^2$ and interpretation of $M^\alpha$, would be to make our universe still roughly two-dimensional
(a bit like real-world galaxies), just adding one extra dimension of our space that extrudes the tessellation and is not rotated.
\end{itemize}

\paragraph{Acknowledgments}
This work has been supported by the National Science Centre, Poland, grant UMO-2019/35/B/ST6/04456.

\def\ext#1{{\it #1}}
\bibliographystyle{plain}
\bibliography{relhell}
\end{document}